\documentclass[journal=jacsat,manuscript=article]{achemso}
\setkeys{acs}{articletitle = true}
\setkeys{acs}{etalmode=truncate,maxauthors=0}

\usepackage[version=3]{mhchem} 
\usepackage{bm}
\usepackage{tabu}
\usepackage{esvect}
\usepackage{natbib}
\usepackage{pdfpages}

\newcommand{\sml}[1]{ \scriptscriptstyle #1 }

\newcommand{\ket}[1]{|#1\rangle}
\newcommand{\Braket}[1]{\langle #1\rangle}

\usepackage{xcolor,colortbl}
\usepackage{color}

\def \mup {m^{\sml \uparrow}}
\def \mdown {m^{\sml \downarrow}}
\def \nup {n^{\sml \uparrow}}
\def \ndown {n^{\sml \downarrow}}

\author{Bayileyegn A. Abate}
\affiliation[{Science of Advanced Materials, Central Michigan University, Mount Pleasant, MI, 48859, USA}]
{Science of Advanced Materials, Central Michigan University, Mount Pleasant, MI, 48859, USA}
\altaffiliation{Both authors contributed equally to this work.}
\author{Rajendra P. Joshi}
\affiliation[{Department of Physics, Central Michigan University, Mount Pleasant, MI, 48859, USA}]
{Department of Physics, Central Michigan University, Mount Pleasant, MI, 48859, USA}
\affiliation[{Science of Advanced Materials, Central Michigan University, Mount Pleasant, MI, 48859, USA}]
{Science of Advanced Materials, Central Michigan University, Mount Pleasant, MI, 48859, USA}
\altaffiliation{Both authors contributed equally to this work.}
\author{Juan E. Peralta}
\email{peral1j@cmich.edu}
\affiliation[{Department of Physics, Central Michigan University, Mount Pleasant, MI, 48859, USA}]
{Department of Physics, Central Michigan University, Mount Pleasant, MI, 48859, USA}
\affiliation[{Science of Advanced Materials, Central Michigan University, Mount Pleasant, MI, 48859, USA}]
{Science of Advanced Materials, Central Michigan University, Mount Pleasant, MI, 48859, USA}

\title{Local Noncollinear Spin Analysis}

\abbreviations{IR,NMR,UV}

\begin{document}

\begin{abstract}

In this work, we generalize the local spin analysis of Clark and Davidson [J.
Chem. Phys. 115(16), 7382 (2001)] for the partitioning of the expectation value
of the molecular spin square operator, $\langle \hat{\bf S}^2 \rangle$,  into atomic contributions, 
$\langle \hat{\bf S}_A \cdot \hat{\bf S}_B  \rangle$,  to the noncollinear spin case in the
framework of density functional theory (DFT). We derive the working equations and we 
show applications to the analysis of
the noncollinear spin solutions of typical spin-frustrated systems and to the
calculation of magnetic exchange couplings. In the former case, we employ the
triangular H$_3$He$_3$ test molecule and a Mn$_3$ complex to show that the local spin
analysis provides additional information that complements 
the standard one-particle spin population analysis. For the calculation of magnetic exchange
couplings, $J_{AB}$, we employ the local spin partitioning to extract $\langle \hat{\bf S}_A \cdot \hat{\bf S}_B  \rangle$ as a
function of the interatomic spin orientation given by the angle  $\theta$. This, combined with
the dependence of the electronic energy with $\theta$, provides a
methodology to extract  $J_{AB}$ from DFT calculations that, in contrast to conventional energy differences based methods, 
does not require the use of {\em ad-hoc} $S_A$ and $S_B$ values.


\end{abstract}


\pagebreak

\section{Introduction}

Molecular magnets, spin-glasses, and topologically frustrated
anti-ferromagnets are representative
examples of materials exhibiting noncollinear magnetism, where the spins may be
disordered and the direction of the magnetization density varies in
space.\cite{N-collinear,Peralta:2007a, Frustated-M1,Spin-glass} 
 Noncollinearity
in the spin direction usually  originates in the  geometric frustration of anti-ferromagnetic 
interactions, magnetic anisotropy effects, or are induced for particular device
applications.\cite{Origin-NC} 
Although electronic structure methods needed to deal with  such systems 
naturally involve the use of multi-determinant wave functions, 
the typical size and complexity of these systems prohibit the use of
multi-reference wave function methods, and practical calculations are limited to
single-determinant methods.\cite{Podewitz:2008a, Cramer:2009a} 
Over the years,
density-functional theory (DFT) has become one of the most successful and
widely used computational tools for electronic structure theory of complex
chemical systems, 
mainly due to the combination of its low computational cost and the
availability of increasingly accurate approximations to the
exchange-correlation energy.~\cite{Capelle:2006a, Jensen:2007a, Szabo:1996a,
Cramer:2009a, Jacob:2012a} 
Within the DFT formalism, a  general description of the spin degree of freedom
can be realized  by allowing  noncollinear spin magnetization.\cite{Kubler:1988a,Peralta:2004a, Peralta:2007a, Luo:2012a, Scalmanni:2012a, Bulik:2013a}
This generalization, refered to as noncollinear spin DFT, helped to gain insight  into the underlying physics of  materials properties
and chemical processes involving magnetic systems.\cite{Kubler:1988a, Peralta:2004a, PhysRevB.74.140405, Peralta:2007a, Luo:2012a, Scalmanni:2012a, Bulik:2013a, spin-dynamics}
Although a wealth of
DFT calculations currently employ the   noncollinear spin formalism,
the analysis of the resulting spin density is limited to the partitioning of the expectation value of the one-particle spin operator, $\Braket{\hat{\bf S}}$.
This  analysis  provides information about the magnitude and direction of the spin magnetization of different atoms or molecular units but lacks 
information about interatomic spin interactions.  

The concept of local spins\cite{Clark:2001a, Reiher:2007a, Mayer:2007a, Podewitz:2008a, Manz:2011a,doi:10.1021/ct200594f} 
is based on the partitioning of the expectation value of the molecular spin square operator, $\Braket{\hat{{\mathbf{S}}}^{2}}$, and 
provides a valuable input to understand the electronic structure of
molecules that  complements the information obtained from one-particle  population
analysis methods.

In Kohn-Sham (KS) DFT, $\Braket{\hat{{\mathbf{S}}}^{2}}$  is calculated in analogy to \emph{ab initio}
wave function methods utilizing the auxiliary KS orbitals.\cite{Schmidt:2008a}
It should be pointed out that, although this is common practice, in DFT the formally correct way to evaluate $\Braket{\hat{{\mathbf{S}}}^{2}}$  is
trhough a functional of the spin-density.\cite{doi:10.1063/1.468585,doi:10.1063/1.2737773,doi:10.1063/1.2978168}
Local spin analyses rely  on the exact decomposition of the 
$\hat{{\mathbf{S}}}$ operator in terms of local projectors.\cite{Clark:2001a, Clark:2003a, Herrmann:2005a, Reiher:2007a, Mayer:2007a, Mayer:2009a, Alcoba:2009a, Ramos-Cordoba:2012a, Ramos-Cordoba:2012b,  Ramos-Cordoba:2013a} 
The general framework originally proposed by Clark and
Davidson~\cite{Clark:2001a} employs Hermitian one-electron projection operators to decompose
$\Braket{\hat{{\mathbf{S}}}^{2}}$ into one- and two-center contributions, $\Braket{\hat{\mathbf{S}}^{2}_{A}}$ and
$\Braket{\hat{{\mathbf{S}}}_{A}\cdot\hat{{\mathbf{S}}}_{B}}_{A\neq{B}}$,    
respectively. The pioneer method of Clark and Davidson  has been used in several contexts
to characterize local collinear spins for various systems, including organic radicals and 
transition-metal complexes.\cite{Herrmann:2005a, Podewitz:2008a}. Other alternative 
decomposition schemes have been proposed in the literature.\cite{Mayer:2007a,Ramos-Cordoba:2012b} 
In this work, we extend the local spin analysis of Clark and Davidson to  the general noncollinear
spin case.  As proof-of-concept, we apply the local spin analysis to the 
triangular H$_3$He$_3$ test molecule and a Mn$_3$ complex, both showing noncollinear spin solutions originated by geometrically frustrated anti-ferromagnetic interactions.
We also show that the   local noncollinear spin analysis can be used as a tool to  extract magnetic
exchange coupling parameters from a unique  single-reference high spin state without  
 {\em ad-hoc} assumptions about nominal $S_A$ and $S_B$ spin values.


%


\section{\label{sec:level2} Theory }
In noncollinear spin DFT, spin noncollinearity  is introduced through two-component Kohn-Sham complex  spinors 
 \begin{eqnarray}
 	\label{eq:spinors}
        \psi_{i}(\mathbf{r}) &=& \left(
        \begin{array}{c}
        \psi_{i}^{\uparrow}(\mathbf{r}) \\[1.2ex]
        \psi_{i}^{\downarrow}(\mathbf{r})
        \end{array} \right),
\end{eqnarray}       
 where the spatial orbitals, $\psi_{i}^{\uparrow}(\mathbf{r})$ and $ \psi_{i}^{\downarrow}(\mathbf{r})$ are expanded in terms of atomic orbitals,
 \begin{eqnarray}
  \psi_{i}^{\sigma}({\mathbf{r}}) &=& \sum_{\nu}c_{\nu i}^{\sigma}\phi_{\nu}({\mathbf{r}}) \qquad (\sigma = \uparrow, \downarrow).
\end{eqnarray}
For the purpose of this work, it is convenient to use the one-electron density matrix
\begin{eqnarray}
D_{\mu\nu} = \sum_{i \in  occ} 
 \left( \begin{array}{cc}
c_{\mu i }^\uparrow c_{\nu i }^{\uparrow {\ast}}  & c_{\mu i }^\uparrow c_{\nu i }^{\downarrow {\ast}}  \\
c_{\mu i }^\downarrow c_{\nu i }^{\uparrow {\ast}}  &   c_{\mu i }^\downarrow c_{\nu i }^{\downarrow {\ast}}   \end{array} \right) 
=  \left( \begin{array}{cc}  
\mathbf{D}^{\uparrow\uparrow}_{\mu\nu} & \mathbf{D}^{\uparrow\downarrow}_{\mu\nu}\\
\mathbf{D}^{\downarrow\uparrow}_{\mu\nu} & \mathbf{D}^{\downarrow\downarrow}_{\mu\nu}  \end{array} \right)     \,,
\label{Eq:GP}
\end{eqnarray}
\noindent where $\mathbf{D}^{\sigma\sigma'}_{\mu\nu} $ are the four spin blocks of the complex density matrix  used in this local noncollinear  spin partitioning.

The local projection operator associated with atom $A$, $ \hat{P}_{A}$ is used to project
the contribution of atom $A$ from the total molecular spin. The total and local
spin operators can be written as \begin{eqnarray}
	 \hat{{\mathbf{S}}} &=& \sum_{i}\hat{{\mathbf{S}}}(i), \, \text{and} \\
	\hat{{\mathbf{S}}}_{A} &=& \sum_{i}\hat{{\mathbf{S}}}(i)\hat{P}_{A}(i).
	\end{eqnarray}
Using this definition of $\hat{{\mathbf{S}}}_{A}$, the square of the total spin operator becomes
	\begin{eqnarray}
	\hat{{\mathbf{S}}}^{2} &=& \sum_{A,B}\hat{{\mathbf{S}}}_{A}\cdot\hat{{\mathbf{S}}}_{B},
	\end{eqnarray}
which can be expanded in terms of local projection operators using
\begin{eqnarray}
\label{eq:sab}
\hat{{\mathbf{S}}}_{A}\cdot\hat{{\mathbf{S}}}_{B} &=& \sum_{ij}\hat{{\mathbf{S}}}(i)\hat{P}_{A}(i)\cdot\hat{{\mathbf{S}}}(j)\hat{P}_{B}(j)\nonumber \\[2ex]
&=& \delta_{AB}  \sum_{i}\hat{P}_{A}(i) \hat{S}^2(i)
+ \frac{1}{2} \sum_{i,j}\hat{{\mathbf{S}}}(i)\cdot\hat{{\mathbf{S}}}(j)[\hat{P}_{A}(i)\hat{P}_{B}(j) + \hat{P}_{B}(i)\hat{P}_{A}(j)]
\end{eqnarray}
In Eq.~(\ref{eq:sab}), the first and second terms on the right-hand side
represent one- and two-electron operators, respectively. For a single-reference
method, such as Hartree-Fock or  Kohn-Sham DFT,\cite{Jensen:2007a, Szabo:1996a}
the expectation value $\Braket{\hat{{\mathbf{S}}}_{A}\cdot\hat{{\mathbf{S}}}_{B}}$ is
given by \begin{eqnarray}
\Braket{\hat{{\mathbf{S}}}_{A}\cdot\hat{{\mathbf{S}}}_{B}} 
& = & \frac{3}{4} \delta_{AB}  \sum_{m}\Braket{m|\hat{P}_{A}|m} + \frac{1}{2}\sum_{m,n}\Braket{mn|\hat{{\mathbf{S}}}(1)\cdot\hat{{\mathbf{S}}}(2)[\hat{P}_{A}\hat{P}_{B} + \hat{P}_{B}\hat{P}_{A} ]|mn} \nonumber \\[2ex]
&& - \frac{1}{2}\sum_{m,n}\Braket{mn|\hat{{\mathbf{S}}}(1)\cdot\hat{{\mathbf{S}}}(2)[\hat{P}_{A}\hat{P}_{B} + \hat{P}_{B}\hat{P}_{A} ]|nm},
\end{eqnarray}
where $m$ and $n$ refer to two-component spinors, Eq.~(\ref{eq:spinors}). Using
the fact that spin and projection operators commute, and working the algebra
for the two-component spinors, the expectation
value $\Braket{\hat{{\mathbf{S}}}_{A}\cdot\hat{{\mathbf{S}}}_{B}}$ in terms of
the projection operators, $\hat{P}_{A}$ and $\hat{P}_{B}$, can be cast as
\begin{eqnarray}
\label{eq:finaleq1}
\Braket{\hat{{\mathbf{S}}}_{A}\cdot\hat{{\mathbf{S}}}_{B}} & = &
\frac{3}{4}\delta_{AB} \sum_{m} \left[\Braket{\mup|\hat{P}_{A}|\mup} + \Braket{\mdown|\hat{P}_{A}|\mdown} \right] + \nonumber \\[2ex]
&& \sum_{m,n} \bigg[ \frac{1}{4} \Braket{\mup|\hat{P}_{A}|\mup}\Braket{\nup|\hat{P}_{B}|\nup} +
\frac{1}{4} \Braket{\mdown|\hat{P}_{A}|\mdown} \Braket{\ndown|\hat{P}_{B}|\ndown} \nonumber \\[2ex]
&& - \frac{1}{4} \Braket{\mup|\hat{P}_{A}|\nup}\Braket{\nup|\hat{P}_{B}|\mup} -
\frac{1}{4}\Braket{\mdown|\hat{P}_{A}|\ndown} \Braket{\ndown|\hat{P}_{B}|\mdown} \nonumber \\[2ex]
&& - \frac{1}{4} \Braket{\mup|\hat{P}_{A}|\mup} \Braket{\ndown|\hat{P}_{B}|\ndown} -
\frac{1}{4} \Braket{\mdown|\hat{P}_{A}|\mdown} \Braket{\nup|\hat{P}_{B}|\nup} \nonumber \\[2ex]
&& - \frac{1}{2} \Braket{\mup|\hat{P}_{A}|\nup} \Braket{\ndown|\hat{P}_{B}|\mdown} -
\frac{1}{2} \Braket{\mdown|\hat{P}_{A}|\ndown} \Braket{\nup|\hat{P}_{B}|\mup} \nonumber \\[2ex]
&& + \frac{1}{2}\Braket{\mup|\hat{P}_{A}|\mdown} \Braket{\ndown|\hat{P}_{B}|\nup} +
\frac{1}{2} \Braket{\mdown|\hat{P}_{A}|\mup} \Braket{\nup|\hat{P}_{B}|\ndown} \nonumber \\[2ex]
&& + \frac{1}{4}\Braket{\mup|\hat{P}_{A}|\ndown} \Braket{\ndown|\hat{P}_{B}|\mup} +
\frac{1}{4} \Braket{\mdown|\hat{P}_{A}|\nup} \Braket{\nup|\hat{P}_{B}|\mdown} \bigg] \,.
\end{eqnarray}
In Eq.~\ref{eq:finaleq1}, $\ket{  m^{\sigma}}$ represents the $\sigma$ space orbital associated with the spinor $\ket{\psi_m}$, so that
 $\Braket{ m^{\sigma}|\hat{P}_X |n^{\sigma'}}= \int \!  d^3\!  r \psi_{m}^{\sigma * } (\mathbf{r})  \hat{P}_X  \psi_{n}^{\sigma'}(\mathbf{r})  $.
For the purposes of this decomposition procedure, there are certain 
conditions that the local projection operators are required to fulfill. First, the
projection operators should be idempotent and orthogonal. Second, the sum of
all projectors must sum up to the identity operator, $\sum_A \hat{P}_A = \hat{I} $ .\cite{Herrmann:2005a, Reiher:2007a} 
Using these conditions, the final working expression for $ \Braket{\hat{{\mathbf{S}}}_{A}\cdot\hat{{\mathbf{S}}}_{B}}$ is
\begin{eqnarray}
\label{eq:finaleq2}
\Braket{\hat{{\mathbf{S}}}_{A}\cdot\hat{{\mathbf{S}}}_{B}} & = & 
\frac{3}{4} \delta_{AB}  \sum_{\mu\in{A}} \Big[ {\mathbf{P}}_{\mu\mu}^{\uparrow\uparrow} + 
{\mathbf{P}}_{\mu\mu}^{\downarrow\downarrow}  \Big]  + 
\sum_{\mu\in{A},\nu\in{B}} 
\Big[
      \frac{1}{4}      {\mathbf{P}}_{\mu\mu}^{\uparrow\uparrow}    {\mathbf{P}}_{\nu\nu}^{\uparrow\uparrow} + 
      \frac{1}{4}      {\mathbf{P}}_{\mu\mu}^{\downarrow\downarrow}{\mathbf{P}}_{\nu\nu}^{\downarrow\downarrow}\nonumber \\[2ex]
&& -  \frac{1}{4}      {\mathbf{P}}_{\mu\nu}^{\uparrow\uparrow}    {\mathbf{P}}_{\nu\mu}^{\uparrow\uparrow} - 
      \frac{1}{4}      {\mathbf{P}}_{\mu\nu}^{\downarrow\downarrow}{\mathbf{P}}_{\nu\mu}^{\downarrow\downarrow} - 
      \frac{1}{4}      {\mathbf{P}}_{\mu\mu}^{\uparrow\uparrow}    {\mathbf{P}}_{\nu\nu}^{\downarrow\downarrow} \nonumber  - 
      \frac{1}{4}      {\mathbf{P}}_{\mu\mu}^{\downarrow\downarrow}{\mathbf{P}}_{\nu\nu}^{\uparrow\uparrow} - 
      \frac{1}{2}      {\mathbf{P}}_{\mu\nu}^{\uparrow\uparrow}    {\mathbf{P}}_{\nu\mu}^{\downarrow\downarrow}  \\[2ex]
&& -  \frac{1}{2}      {\mathbf{P}}_{\mu\nu}^{\downarrow\downarrow}{\mathbf{P}}_{\nu\mu}^{\uparrow\uparrow} +
      \frac{1}{2}      {\mathbf{P}}_{\mu\mu}^{\uparrow\downarrow}  {\mathbf{P}}_{\nu\nu}^{\downarrow\uparrow} + 
      \frac{1}{2}      {\mathbf{P}}_{\mu\mu}^{\downarrow\uparrow}  {\mathbf{P}}_{\nu\nu}^{\uparrow\downarrow} + 
      \frac{1}{4}      {\mathbf{P}}_{\mu\nu}^{\uparrow\downarrow}  {\mathbf{P}}_{\nu\mu}^{\downarrow\uparrow} +
      \frac{1}{4}      {\mathbf{P}}_{\mu\nu}^{\downarrow\uparrow}  {\mathbf{P}}_{\nu\mu}^{\uparrow\downarrow} 
\Big] \,, 
\end{eqnarray}
where $ {\mathbf{P}}_{\eta\zeta}^{\sigma\sigma'} $ are the {\em projected} one-electron density matrix elements  
(for a detailed derivation of Eq.~\ref{eq:finaleq2} please see the Supporting Information)
%
For the collinear spin case (assuming spin-polarization in the $z$ direction),  the contributions from the cross-terms in
the generalized spin-density matrices are all zero, and hence the last four terms in Eq.~\ref{eq:finaleq2} 
vanish,  giving an expression that is equivalent to Eq.~(16) in the paper of Hermann {\em et al}.\cite{Herrmann:2005a}.
For the purpose of implementing the local spin analysis,  it is convenient to attempt to compact the notation in Eq.~\ref{eq:finaleq2}. To this end, we define 
the vector $\vv{\bf{P}}_{\mu\nu} $ of Cartesian components  
$\mathbf{P}_{\mu\nu}^{x} =\mathbf{P}_{\mu\nu}^{\uparrow\downarrow} + \mathbf{P}_{\mu\nu}^{\downarrow\uparrow} $,
$\mathbf{P}_{\mu\nu}^{y} =i( \mathbf{P}_{\mu\nu}^{\uparrow\downarrow} - \mathbf{P}_{\mu\nu}^{\downarrow\uparrow}) $, 
$\mathbf{P}_{\mu\nu}^{z} =\mathbf{P}_{\mu\nu}^{\uparrow\uparrow} - \mathbf{P}_{\mu\nu}^{\downarrow\downarrow} $, and
the scalar
$\mathbf{P}_{\mu\nu} =\mathbf{P}_{\mu\nu}^{\uparrow\uparrow} + \mathbf{P}_{\mu\nu}^{\downarrow\downarrow} $. 
Using these matrices, Eq.~\ref{eq:finaleq2} can be reduced to
\begin{eqnarray} 
\label{eq:finalsimpli}
\Braket{\hat{{\mathbf{S}}}_{A}\cdot\hat{{\mathbf{S}}}_{B}} = \frac{3}{4}(\mathbf{P}_{AA} \delta_{AB} - \frac{1}{2} \mathbf{P}_{AB}  \mathbf{P}_{BA} )
+\frac{1}{4}( \vv{\bf{P}}_{AA} \cdot \vv{\bf{P}}_{BB}  + \frac{1}{2} \vv{\bf{P}}_{AB} \cdot \vv{\bf{P}}_{BA}    ), 
\end{eqnarray}
where we have used the fact that the 
generalized density matrix $\mathbf{P}_{\mu\nu}^{\sigma\sigma}$  is Hermitian, and the indexes $A$ and $B$ imply the summations $\sum_{\mu\in{A}}$ and
 $\sum_{\nu\in{B}}$. It is interesting to note that since  $\vv{\bf{P}}_{\mu\nu} $  transforms as a vector, Eq.~\ref{eq:finalsimpli} trivially reflects 
the rotational invariance of the decomposition of $\Braket{\hat{{\mathbf{S}}}^{2}}$.
Eqs.~\ref{eq:finaleq2} and \ref{eq:finalsimpli} are  general and the main result of this work.
Out of several alternatives available in the literature for the choice of the projector,\cite{Herrmann:2005a,Mayer:2007a,Ramos-Cordoba:2012b} 
L\"{o}wdin projection operators, as originally employed in the work of Clark and Davidson\cite{Clark:2001a} are among the most widely used and simplest to implement.\cite{Lowdin:1950a}
Hence, in this work, the implementation has been carried out using L\"{o}wdin projectors using 
${\mathbf{P}}^{\sigma\sigma'} = {\mathbf{S}}^{1/2}{\mathbf{D}}^{\sigma\sigma'}{\mathbf{S}}^{1/2}$, where $\mathbf{S} $ is the 
overlap matrix.

\section{Computational Details}

The $\Braket{\hat{{\mathbf{S}}}^{2}}$ decomposition scheme presented in the previous section has been implemented in an in-house 
Gaussian development version.\cite{gdvh32}  Five  density functional approximations  in combination  with the triple-zeta split valence basis
set with polarization~\cite{Schafer:1994a}, TZVP, were  employed to perform the
noncollinear spin calculations on the two benchmark systems: The triangular
H$_{3}$He$_{3}$ and the polynuclear oxomanganese complex
[(Mn$^{{{4}}}$)$_3$O$_4$L$_4$(H$_2O$)] ([Mn$_{3}$] complex).\cite{Luo:2012a} 
Both systems can be considered as antiferromagnetically coupled  spin trimers 
with  frustrated noncollinear spin configurations. 
We included in our tests the local spin density approximation (LSDA) built as 
Slater  exchange and the parametrization of Wosko, Wilk, and Nusair\cite{Slater1974,vosko80} for correlation,
two functionals from the generalized gradient approximation:  
the Perdew, Burke, and Ernzerhof (PBE) functional\cite{PBE_1} and the Becke's 1988 exchange  plus Perdew's 1996 correlation (BP86),\cite{PhysRevA.38.3098,PhysRevB.33.8822}
and two  representative hybrid functionals:
PBEh\cite{PBE0a} (PBE hybrid, also refer to as  PBE1PBE\cite{PBE0b} and  PBE0\cite{PBE0c} in the literature) and B3LYP.\cite{PhysRevA.38.3098,lyp,Becke_JCP_1993_B,B3LYP,HERTWIG1997345} 

In order to  find these  noncollinear spin configurations in the self-consistent
solutions of the KS equations, it is important to start from a suitable initial
guess. For this work, the initial guesses were thus generated by pre-conditioning the
electron  spin density  in a noncollinear configuration using a constraint that imposes 
local-spin moments in pre-selected directions. These constraints are introduced
{\em via} Lagrange multipliers. 
To this end, we write the  local magnetization for 
atom $A$ as \begin{eqnarray}
{\mathbf m}^A =
\sum_{\mu,\nu} {\mathcal W}^A_{\mu\nu} {\mathbf D}_{\mu\nu} \,,
\label{eq:S-pop}
\end{eqnarray}
where $ {\mathcal W}^A$ is a local projector,    ${\mathbf D}_{\mu\nu} $ is the spin-density matrix vector with
Cartesian components
$D^x_{\mu\nu}   =  D^{\uparrow \downarrow}_{\mu\nu}  + D^{ \downarrow
\uparrow}_{\mu\nu} $, 
$D^y_{\mu\nu}    =   i ( D^{\uparrow \downarrow}_{\mu\nu} -
D^{\downarrow \uparrow}_{\mu\nu} )$, 
and
$D^z_{\mu\nu}    =   D^{\uparrow \uparrow}_{\mu\nu}  - D^{\downarrow \downarrow}_{\mu\nu}$.
In Eq.~(\ref{eq:S-pop}), ${\mathcal W}^A$ is defined from the
L\"{o}wdin  partitioning as
\begin{eqnarray}
{\mathcal W}^A_{\mu\nu} =  \sum_{\lambda \in A}
({\mathbf S}^{1/2})_{\mu\lambda} ({\mathbf S}^{1/2})_{\lambda\nu}\,,
\end{eqnarray}
where $\mathbf S $ is the AO overlap matrix.
For each atom $A$ for which the  local magnetization $ {\mathbf{m}}^A$ is to be
constrained in a direction ${\mathbf e}^A$,
an aditional term ${\mathbf h}^A$  is included in ${\mathbf H}_{KS}$,
\begin{eqnarray}
{\mathbf h}^{A} = {\mathcal W}^A  {\bm\lambda}^A \cdot( {\bm \sigma} \times {\mathbf e}^A )\,,
\end{eqnarray}
where ${\mathbf e}^A$ is a unit vector to which ${\mathbf m}^A$ is constraint to
be parallel to,
${\bm \lambda}^A$ is a Lagrange multiplier vector and ${\bm \sigma}$ are
the Pauli matrices.
For simplicity, and for the purpose of generating an initial guess, in this work  we
use a fixed value for ${\bm\lambda}^A=0.5$. For the magnetic exchange coupling application in the next Section), 
two multipliers, $\bm\lambda^A$ and $\bm\lambda^B$ 
are introduced (one for each atom $A$ and $B$) and the values of  $\bm\lambda^A$  and $\bm\lambda_B$ 
are variationally optimized to minimize the KS energy during 
the self-consistent iterations.
The total and local spin values reported in this work are in atomic units, and no 
symmetry was imposed (keyword NoSymm in Gaussian) in all  the calculations.

\section{\label{sec:results}Results and Discussion}

\subsection{Spin Trimers}
\label{sec:trimers}

We have characterized a noncollinear spin configuration  of the triangular 
H$_{3}$He$_{3}$ test molecule, which is equivalent to a C$_{3h}$  spin trimer with
S=1/2 at each center. Figure~1(a) shows the H$_{3}$He$_{3}$ structure with the
hydrogens at the vertices of the  triangle of side 3~\AA, and the He
atoms on the midpoint of each side, giving place to an effective antiferromagnetic superexchange coupling between the H atoms. 
The ground state of
H$_{3}$He$_{3}$ is composed of two degenerate doublets (S=1/2) that can be
described by four degenerate spin wave functions. Four different compromised
noncollinear spin arrangements can be obtained from linear combinations of
these degenerate spin wave functions.\cite{Dai:2004a} The scheme in
Figure~1(a) shows one of these noncollinear spin arrangements obtained in the
present calculation, where 
all the spin moments are of equal magnitude and lie on the plane of the molecule, and 
the spin-spin angle between neighboring spins of
120$^\circ$ (it should be mentioned that in the absence of spin-orbit, as it is the case here,
the total magnetization can be rigidly rotated without changing the total electronic energy).
Table {1} summarizes the total and local spins
calculated using  different density functional approximations. It has been noted 
previously that the local spins obtained using the
projection operators scheme show basis-set and functional dependence.\cite{Herrmann:2005a}  
For this particular 
system, however, the functional dependence of the local spins is very small in this case, and 
in all cases the local spin at each H center is reasonably close to the ideal
value of 0.75 for a spin-1/2 center, as expected,  due to the localized nature of the magnetic moments.
%
%

\begin{figure}\centering
  \includegraphics[width=0.75\textwidth]{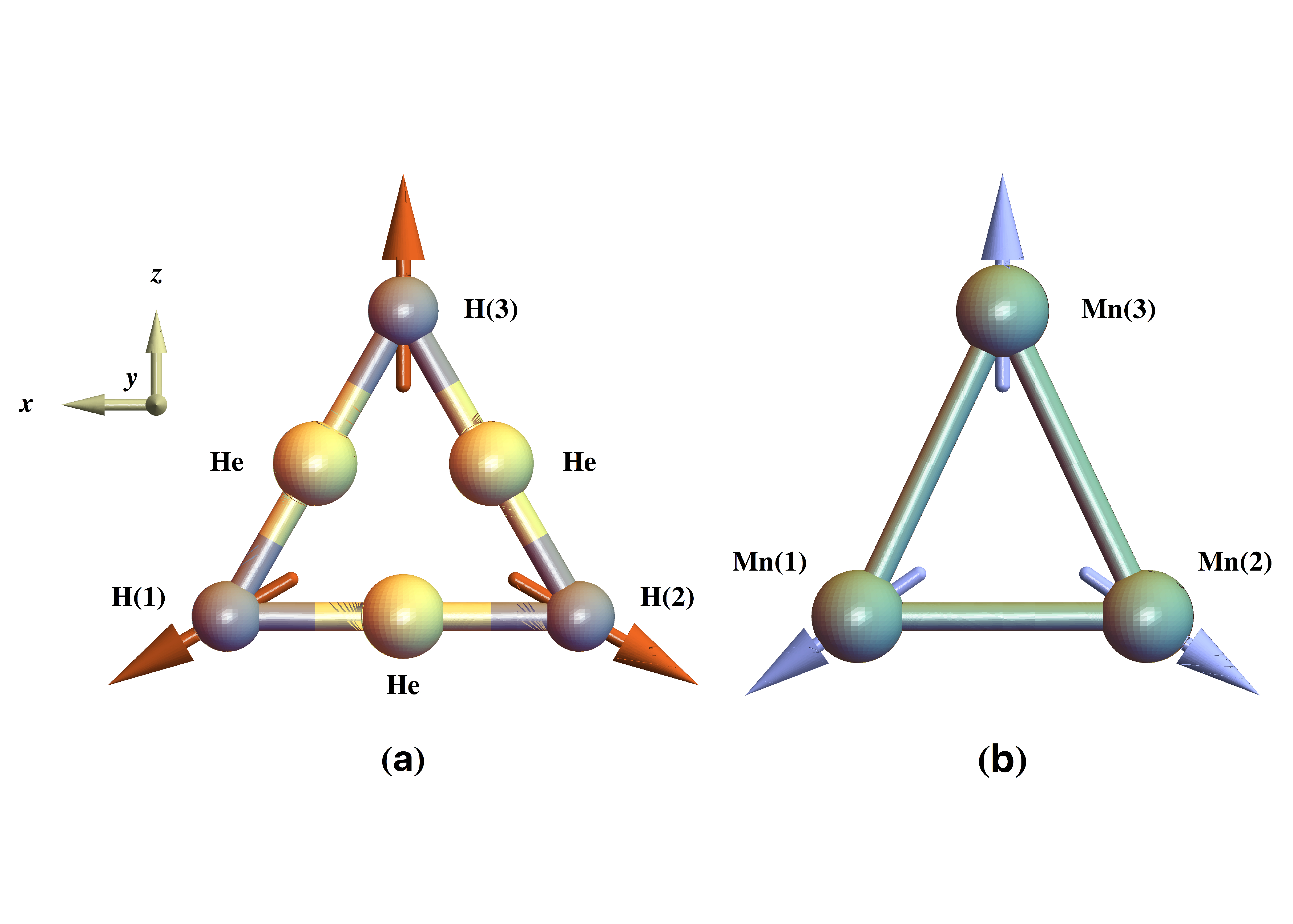}
\caption{Schematic representation of the  H$_{3}$He$_{3}$ molecule (a) and [Mn$_3$] core (b) with their respective noncollinear spin configurations shown as arrows. }
\end{figure}

%
%
  \begin{table}[htp]
\caption{Calculated total and local spin values for H$_{3}$He$_{3}$.  $\Braket{\hat{{\mathbf{S}}}_{\text{H}}\cdot\hat{{\mathbf{S}}}_{\text{H$'$}}}$  refers to the off-diagonal terms in the local spin analysis.    }
\begin{center}
\begin{tabular}{|l|c|c|c|c|}
 \hline
 \hline
Method & $\Braket{\hat{{\mathbf{S}}}^{2}}$	& $|\Braket{\hat{\mathbf{S}}_{\text{H}}}|$	& $\Braket{\hat{{\mathbf{S}}}_{\text{H}}^{2}}$	& $\Braket{\hat{{\mathbf{S}}}_{\text{H}}\cdot\hat{{\mathbf{S}}}_{\text{H$'$}}}$		\\ 
 \hline
LSDA	& 1.45	& 0.45	& 0.71	& $-$0.11	\\
 PBE	& 1.48	& 0.46	& 0.72	& $-$0.11	\\
 BP86	& 1.47	& 0.46	& 0.72	& $-$0.11	\\
 B3LYP	& 1.48	& 0.46	& 0.72	& $-$0.11	\\	
 PBEh	& 1.48	&  0.47	& 0.73	& $-$0.11	\\
 \hline
\hline
\end{tabular}
\end{center}
\end{table}%

For the [Mn$_{3}$] complex considered in our test calculations, we have used
the relaxed  structure from the work of Luo {\em et al}.~\cite{Luo:2012a} to calculate
local spins for both the noncollinear and ncollinear spin cases (Supporting Information). Considering the C$_{2h}$
core structure of this complex, the three Mn are located at the vertices of an
isosceles triangle, shown in Figure 1(b). 
In the  actual structure of the complex, Mn(1) and Mn(2) are coupled with Mn(3)
by a O-Mn-O linkage. As it can be expected,  most of the molecular spin is
localized at  the three Mn centers. 
The discussion here is focused on the
$\Braket{\hat{{\mathbf{S}}}_{A}\cdot\hat{{\mathbf{S}}}_{B}}$ values pertinent
to the three Mn centers. Table 2 summarizes the total and local noncollinear
spins calculated using  different DFT approximations. Mn(1) and Mn(2), which
are closer to each other than to Mn(3), are equivalent in this noncollinear solution and have the same local
spin values. 
From the local spin values $\Braket{\hat{{\mathbf{S}}}_\text{Mn}^{2}}$,
it can be easily confirmed that the Mn centers are all in the local high-spin
state. 
As pointed out previously,\cite{Clark:2001a,
Herrmann:2005a, Reiher:2007a} even if these values cannot be formally interpreted as
S(S+1), it is  interesting to note that the
$\Braket{\hat{{\mathbf{S}}}^{2}_\text{Mn(3)}}$, except for PBEh, are
close to the ideal value of 3.75 for a spin-3/2 center. 
The corresponding values for the other two centers, Mn(1) and Mn(2), are indeed noticeably different
from the ideal S(S+1). 
It can also be observed in Table~2  that both, the
total and local values $\Braket{\hat{{\mathbf{S}}}^{2}_\text{Mn}}$ , show a systematic increase with the incorporation of  Hartree-Fock exchange, as expected from the higer electron localization of the $d$ electrons. 
The off-diagonal terms
$\Braket{\hat{{\mathbf{S}}}_{A}\cdot\hat{{\mathbf{S}}}_{B}}_{A\neq{B}}$, which
indicate the presence and nature of the spin-spin coupling between the local
spins, are all negative, corresponding to an antiferromagnetic arrangement  between
neighboring spins. The different functionals tested in this work give very similar values for the
spin-spin angles ($\theta_{12}\approx$150$^{\circ}$ and  $\theta_{13}\approx\theta_{23}\approx$105$^{\circ}$, as 
calculated from $\hat{\mathbf{S}}_\text{Mn(3)}$ (not shown here); however, the off-diagonal local spin values reflect a decrease of the 
antiferromagnetic Mn(1)--Mn(2) interaction and an increase of the Mn(1,2)--Mn(3) antiferromagnetic 
interaction with hybrid functionals. 
This can be interpreted as originated in an decreased metal-ligand interaction, as 
quantified by $\Braket{\hat{{\mathbf{S}}}_{\text{Mn}}\cdot\hat{{\mathbf{S}}}_{\text{L}}}$ 
 in Table 2. It is interesting to note that the  local spin 
at the non-metal centers  $\Braket{\hat{{\text{S}}}_{\text{L}}^2}$, remains almost constant for all DFT approximations 
tested here. 


The calculated total and local
spin values for the collinear spin case corresponding to one of the  broken-symmetry solutions  with magnetization 
$\uparrow$  at Mn(1) and $\downarrow$ at Mn(2) and Mn(3) are summarized in Table~3. 
For this broken-symmetry solution, the calculated $\Braket{\hat{{\mathbf{S}}}^{2}_\text{Mn(1,2)}}$ follow a 
similar  trend as in the noncollinear spin case, while $\Braket{\hat{{\mathbf{S}}}^{2}_\text{Mn(3)}}$ remains 
almost constant for the five funcionals tested here. This can be explained from the  competing 
ferromagnetic $\Braket{\hat{{\mathbf{S}}}_{\text{Mn(2)}}\cdot\hat{{\mathbf{S}}}_{\text{Mn(3)}}}$   
and  antiferromagnetic  $\Braket{\hat{{\mathbf{S}}}_{\text{Mn(1)}}\cdot\hat{{\mathbf{S}}}_{\text{Mn(3)}}}$ interactions. 
As Hartree-Fock exchange is incorporated, the magnetic exchange coupling between Mn centers becomes weaker. 
This effectively increases the magnitude of the metal $\Braket{\hat{{\mathbf{S}}}_{A}\cdot\hat{{\mathbf{S}}}_{B}}$ 
 in the collinear spin case, while in the noncollinear (frustrated)  spin case, this is not 
possible for all metal pairs.

\begin{table}[h]
\caption{Calculated total and local spin values for the [Mn$_{3}$] complex (noncollinear spin case). The indices 
label Mn centers as shown in Fig.~1. $\Braket{\hat{{\mathbf{S}}}_{\text{Mn}}\cdot\hat{{\mathbf{S}}}_{\text{L}}}$  
   stands for the sum of the contributions of all 
three Mn centers and all non-metal atoms, and $\Braket{\hat{{\text{S}}}_{\text{L}}^2}$  for all non-metal  atoms.}
\begin{center}
\begin{tabu}{|l|c|c|c|c|c|c|c|}
 \hline
 \hline
\rowfont{\small}
 Method &
$\Braket{\hat{\text{S}}^{2}}$	& 
$\Braket{\hat{\text{S}}_{\text{1,2}}^{2}}$	& 
$\Braket{\hat{\text{S}}_{\text{3}}^{2}}$	& 
$\Braket{\hat{{\mathbf{S}}}_{\text{1}}\cdot\hat{{\mathbf{S}}}_{\text{2}}}$	& 
$\Braket{\hat{{\mathbf{S}}}_{\text{1,2}}\cdot\hat{{\mathbf{S}}}_{\text{3}}}$ & 
$\Braket{\hat{{\mathbf{S}}}_{\text{Mn}}\cdot\hat{{\mathbf{S}}}_{\text{L}}}$   &  
$\Braket{\hat{{\text{S}}}_{\text{L}}^2}$    \\ 
 \hline
\rowfont{\small}
 LSDA	& 4.85	&  4.32	& 3.64	& $-$1.28	& $-$0.30  & $-$11.48  & 7.81	\\
\rowfont{\small}
 BP86	& 5.26	&  4.50	& 3.72 	& $-$1.38	& $-$0.34  &  $-$11.22 & 7.87	\\ 
\rowfont{\small}
 PBE	& 5.40	&  4.59	& 3.79	& $-$1.43	& $-$0.37  &  $-$11.17 & 7.92	\\
\rowfont{\small}
 B3LYP	& 7.02	&  5.38	& 3.95	& $-$0.79	& $-$0.96  & $-$9.72  & 7.42	\\
\rowfont{\small}
 PBEh	& 8.84	&  5.69	& 4.17	& $-$0.52	& $-$1.17  & $-$8.80  & 7.82	\\
  \hline
 \hline	 
\end{tabu}
\end{center}
\end{table}%
%
%
  \begin{table}[h]
\caption{Calculated total and local spin values for the [Mn$_{3}$] complex (collinear spin case).
 The indices label Mn centers as shown in Fig.~1. 
$\Braket{\hat{{\mathbf{S}}}_{\text{Mn}}\cdot\hat{{\mathbf{S}}}_{\text{L}}}$     
   stands for the sum of the contributions of all
three Mn centers and all non-metal atoms, and $\Braket{\hat{{\text{S}}}_{\text{L}}^2}$  for all non-metal  atoms.  }
\begin{center}
\begin{tabu}{|l|c|c|c|c|c|c|c|c|c|}
\rowfont{\small}
 \hline
 \hline 
 Method & 
$\Braket{\hat{\text{S}}^{2}}$	& 
$\Braket{\hat{\text{S}}_{\text{1}}^{2}}$ &  
$\Braket{\hat{\text{S}}_{\text{2}}^{2}}$  &
$\Braket{\hat{\text{S}}_{\text{3}}^{2}}$	&
$\Braket{\hat{{\mathbf{S}}}_{\text{1}}\cdot\hat{{\mathbf{S}}}_{\text{2}}}$	&
$\Braket{\hat{{\mathbf{S}}}_{\text{1}}\cdot\hat{{\mathbf{S}}}_{\text{3}}}$  & 
$\Braket{\hat{{\mathbf{S}}}_{\text{2}}\cdot\hat{{\mathbf{S}}}_{\text{3}}}$  &	
$\Braket{\hat{{\mathbf{S}}}_{\text{Mn}}\cdot\hat{{\mathbf{S}}}_{\text{L}}}$   &   
$\Braket{\hat{{\text{S}}}_{\text{L}}^2}$  \\ 
\hline
LSDA	& 4.88 	 & 4.69  & 4.14	& 3.53	& $-$1.50	& $-$1.19  & 0.92 & $-$6.38  & 8.83  \\
BP86	& 5.18	 & 4.73  & 4.27 & 3.61 	& $-$1.57  	& $-$1.26  & 1.02 & $-$6.41  & 9.01     \\ 
PBE  	& 5.27   & 4.84	 & 4.32	& 3.69	& $-$1.63   & $-$1.33  & 1.07 & $-$6.42  & 9.06   \\
B3LYP	& 5.68	 & 5.09  & 4.92	& 3.53	& $-$1.97	& $-$1.37  & 1.27 & $-$5.53  & 7.33     \\
PBEh	& 5.93	 & 5.38  & 5.27	& 3.60	& $-$2.23	& $-$1.48  & 1.40 & $-$5.56  & 7.42      \\
\hline
\hline	 
\end{tabu}
\end{center}
\end{table}%


\section{Application: Magnetic Exchange Couplings}

As an application of the local noncollinear  spin analysis developed in this work,
in this Section   we show how it can be used to calculate 
magnetic exchange coupling parameters from a single reference spin state without explicit use of nominal spin values. 
Assuming a pairwise interaction between spins, magnetic
exchange interactions can be  modeled by the  Heisenberg-Dirac-van-Vleck (HDVV) spin
Hamiltonian\cite{Hamiltonian} of the form
\begin{equation}
\label{eq:HDVV}
\hat{H}_{\sml HDVV} = -\sum_{\sml  \langle  i,j \rangle}  J_{ij} \,  \hat{\mathbf S}_i \cdot \hat{\mathbf S}_j \,,
\end{equation}
\noindent  where  $\hat{\mathbf S}_k$ is the local spin
operator on  magnetic  center $k$.  This model spin
Hamiltonian considers the isotropic exchange interaction between local spins associated with localized unpaired
electrons. 
Taking the second-order   derivative of the  expectation value of Eq.~\ref{eq:HDVV} with respect to the inter-spin angle $\theta$
for a dinuclear system with centers $A$ and $B$ in its high-spin (HS) state ,  we can write
\begin{equation}
{\frac{d^2 \Braket{\hat{H}_{\sml HDVV} }    }    { d \theta^2}   {{} }  
 \bigg|_{\sml \theta=0^{\circ}}   }  = -  J_{AB}~ {\frac{d^2  
 \Braket{\hat{{\mathbf{S}}}_{A} \cdot  \hat{{\mathbf{S}}}_{B} }  } 
   { d \theta^2}   \bigg|_{\sml \theta=0^{\circ}} }      ~                 \, . 
 \label{eq:SecDer}
     \end{equation}  
Assuming that the  KS system behaves as the HDVV system upon  differential rotations of the inter-spin angle, we can  
replace $\Braket{\hat{H}_{\sml HDVV} }$ with the KS energy in Eq.~\ref{eq:SecDer} to obtain
\begin{equation}
 J_{AB}  = - \frac{\frac{d^2 {E}_{KS}     }    { d \theta^2}   {{} } ~~  \bigg|_{\sml \theta=0^{\circ}}   }  
 {\frac{d^2   \Braket{\hat{{\mathbf{S}}}_{A} \cdot  \hat{{\mathbf{S}}}_{B} }  }    { d \theta^2}   \bigg|_{\sml \theta=0^{\circ}} }      \,,
\label{eq:J}
\end{equation} 
\noindent where in this case $\theta$ is the angle between the local spin vectors $\hat{{\mathbf{S}}}_{A}$  and $\hat{{\mathbf{S}}}_{B}$ in the KS system. 
Eq.~\ref{eq:J} provides a method to calculate $J_{AB}$ from second derivatives 
with respect to the inter-spin angle
of the total energy and 
$\Braket{\hat{{\mathbf{S}}}_{A} \cdot  \hat{{\mathbf{S}}}_{B} } $.
In previous works we have provided a similar methodology that involved knowledge of the nominal spin values.\cite{JEPVBghfJ,BB-jordan,doi:10.1021/acs.jctc.6b00112} 
 The idea illustrated in this Section provides one 
step further to the determination of $J_{AB}$ without external {\em ad-hoc} parameters.  

To determine the derivatives in the right-hand-side of Eq.~\ref{eq:J}, we
employ   noncollinear constrained DFT as described  in the Computational Details Section and implemented in
Ref.~\citenum{JEPVBghfJ}. The angle between the local magnetic moments is introduced as a constraint in the energy expression and 
the generalized density is fully relaxed subject to that constraint.
We  scan the energy and $\Braket{\hat{{\mathbf{S}}}_{A}
\cdot  \hat{{\mathbf{S}}}_{B} } $  landscapes for small angles and then perform
a numerical fit to obtain the quadratic coefficients in a polynomial expansion.
This strategy is not optimal for production calculations, where a method based on   analytical 
derivatives would be desirable,\cite{BB-jordan} but it serves our
purposes in these  proof-of-concept calculations.
For comparision,  the couplings $J_{AB}$  are also  calculated using the  conventional broken symmetry (BS) energy difference approach,\cite{assumptionmadeExpJ,nonprojected1, Yamaguchi,Illas:2004pb,SP-BSDFT}
\begin{equation}
J_{AB} = \frac{E_{BS}-E_{HS}} {2  S_{A}  S_{B}}  
\label{eq:BS}
\end{equation}
\noindent where $E_{HS}$ and $E_{BS}$ are the energies of the high-spin (HS)  and BS spin solutions,  respectively.

\begin{figure}[htbp]
\begin{center}
\includegraphics[width=1.0\textwidth,  keepaspectratio]{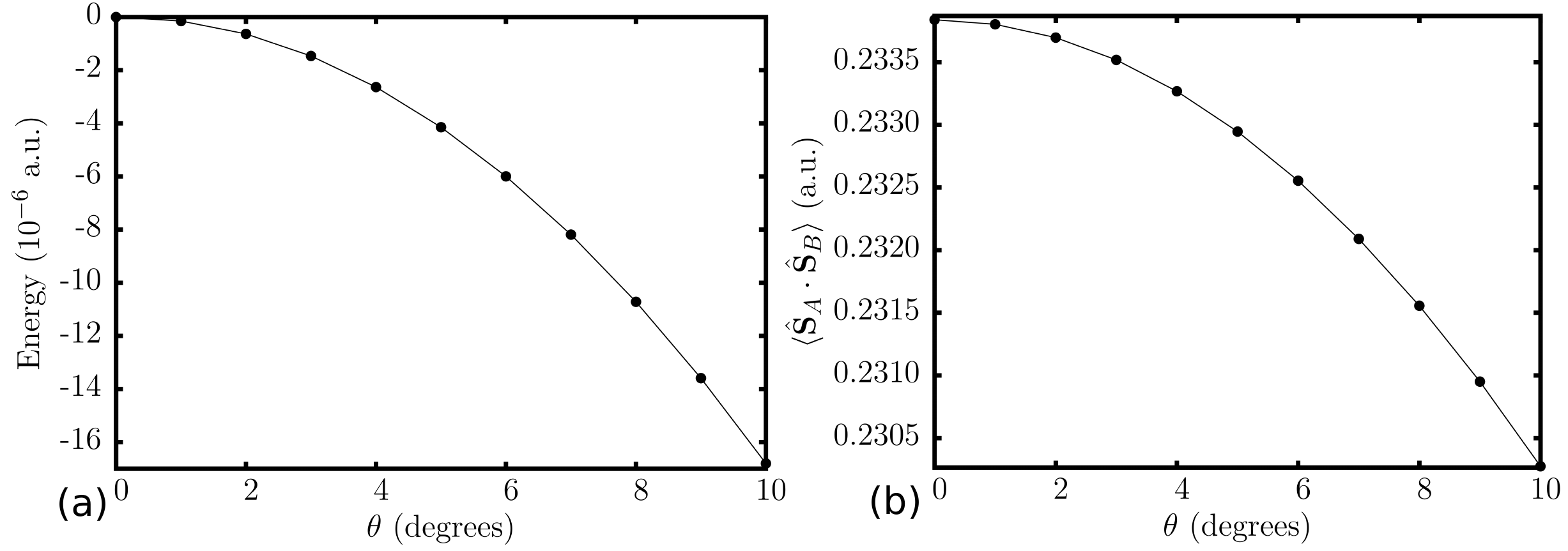}
\caption{KS energy (a) and $\Braket{\hat{{\mathbf{S}}}_\text{H(1)}\cdot\hat{{\mathbf{S}}}_\text{H(2)}}$ (b) variation as a 
function of the angle $\theta$ for the H--He--H molecule (H-He distance of 1.625~\AA).}  
    \label{fig:q}
\end{center}
\end{figure}

To test the proposed methodology, we have selected the H--He--H linear molecule and
the   oxovanadium(IV) dimer [($\mu$-OCH$_3$)VO(ma)]$_2$  (V$_2$ for short; see Supporting Information) as benchmark systems.\cite{V2-exp} For H--He--H, we examine the
performance of the proposed method at different strength of magnetic interactions 
corresponding to two different H--He distances of 1.625~\AA, and 2.0~\AA.
All the calculations in this section were carried out  using
Pople's style split-valence 6-311G** Gaussian basis\cite{6311G,doi:10.1063/1.474865}  
and the B3LYP exchange correlation functional.\cite{PhysRevA.38.3098,lyp,Becke_JCP_1993_B,B3LYP,HERTWIG1997345}
All calculations use a tight convergence criterion of $10^{-8}$~Hartree in the energy.
In Fig.~\ref{fig:q} we show the  KS energy and
$\Braket{\hat{{\mathbf{S}}}_\text{H(1)}\cdot\hat{{\mathbf{S}}}_\text{H(2)}}$ as a function of the 
angle $\theta$ for several different constraint angles about the HS spin state in  H--He--H.
It is clear in this case that  both quantities 
are close to  a quadratic function in  $\theta$, showing that the the contraint DFT method captures the physics of the HDVV model. 
Table~\ref{table:J} shows the exchange couplings $J_{HH}$ obtained from second derivatives (Eq.~\ref{eq:J}) and 
from the BS energy differences approach.
It is worth stressing that, although both methods give close $J$ values, 
the  second derivatives method does not involve 
{\em ad-hoc} parameters, while the BS  method uses $S_H=1/2$ in this case (Eq.~\ref{eq:BS}).
We note  that for the  shortest H--He distance
the relative deviation between both methods is the largerst,  with a 
percentage difference  of about 2.5$\%$. However, this is a somewhat  unrealistic proof-of-concept case, and 
in most cases of practical interest the strength of $J$ is much smaller. Also, it should be pointed out that both methods would exactly agree only 
in the case of a perfectly localized BS spin configuration.

\begin{table}[htp]
\caption{Calculated magnetic exchange couplings (in cm$^{-1}$) for H--He--H molecule obtained from the second derivatives ($J^{SD}$) method, and the BS energy differences 
approach ($J^{BS}$), and percentage deviation. }
\begin{center}
\begin{tabular}{lccc}
 \hline
 \hline  
H-He distance &  $J^{SD}$& $J^{BS}$	 & deviation ($\%$)\\ 
\hline
2.00 \AA	& $-$113.8 ~	&  $-$113.9		&   0.1$\%$	\\
 1.625 \AA	& $-$1051.0~	& $-$1025.6	& 	  2.5$\%$        \\
  \hline
 \hline	 
\end{tabular}
\end{center}
\label{table:J}
\end{table}%

The V$_2$ complex  shows a strong antiferromagnetic 
coupling of about $-$107~cm$^{-1}$, as measured by temperature-dependent magnetic susceptibility experiments.\cite{V2-exp}
For this complex, using the same procedure described above,  we obtained a couplings of $-$179~cm$^{-1}$
with the second derivative approach proposed in this work, and 
 $-$201~cm$^{-1}$ for the BS energy differences approach. 
We found that the small deviation between both methodologies is encouraging, especially considering
that they are not expected to yield identical values for realistic systems such as the V$_2$ complex. 
As mentioned before, from a practical viewpoint a methodology that employs an analytical linear response 
implementation of this method would be desirable to extract both derivatives in Eq.~\ref{eq:J}. Work along this line is in progress.

\section{Concluding Remarks}

In this work we have derived the expressions for the local spin analysis for the case
of a general noncollinear spin single-reference state. This analysis is based on the 
decomposition of the expectation value of the square of the total spin operator and 
utilizes general orthogonal atomic projectors. 
We have also implemented this decomposition  using L\"{o}wdin
projection operators and  showed  its applicability to  characterize the local spins of 
two prototypical cases, H$_{3}$He$_{3}$ and the a [Mn$_{3}$] complex,  where spin noncollinearity 
arises from geometrical frustration of antiferromagnetic interactions.
For both systems, the expected compromised noncollinear spin arrangements are predicted by all the
density functional approximations tested here. While for H$_{3}$He$_{3}$ the local spin
values at each H center are essentially the same for all the functionals,
the [Mn$_{3}$] complex shows a strong dependence on the
functionals used. 
We want to stress  that the local spin analysis requires minimal additional computational resources and can be readily incorporated to 
any electronic structure code with noncollinear spin capabilities.
Besides its potential as an analysis tool, as a practical application we provide a methodology  that utilizes the local spin decomposition for the 
determination of magnetic exchange couplings in magnetic molecules. 
With this method, we evaluate  magnetic exchange couplings from second derivatives of the KS energy and local spin-spin   pair correlation values
$\Braket{\hat{{\mathbf{S}}}_{A} \cdot  \hat{{\mathbf{S}}}_{B} }$ with respect to the interatomic spin angles. 
Our calculations on the benckmark cases H--He--H and  the  oxovanadium(IV) dimer [($\mu$-OCH$_3$)VO(ma)]$_2$ show that our approach to
calculate magnetic exchange couplings using noncollinear local spin analysis
yields couplings  comparable to the ones obtained from energy difference methods. 
However, in contrast to 
traditional energy difference based methods, our  approach does not require any \textit{a priori} knowledge  of  the nominal 
spin values, providing a route to the  blackbox calculation of this property.

\section{Acknowledgements}
This work was supported by the Office of Basic Energy Sciences, US Department of Energy, DE-SC0005027.
JEP thanks Professor Pedro Salvador for helpful comments and suggestions. 

\section{Supporting Information}
The Supporting Information is available free of charge on the ACS Publications website at DOI:. 
Complete derivation of Eq.~10 and 11, and Cartesian corrdinates of the  [Mn$_{3}$] complex and oxovanadium(IV) dimer.
\bibliography{bibliography}

\includepdf[pages=-]{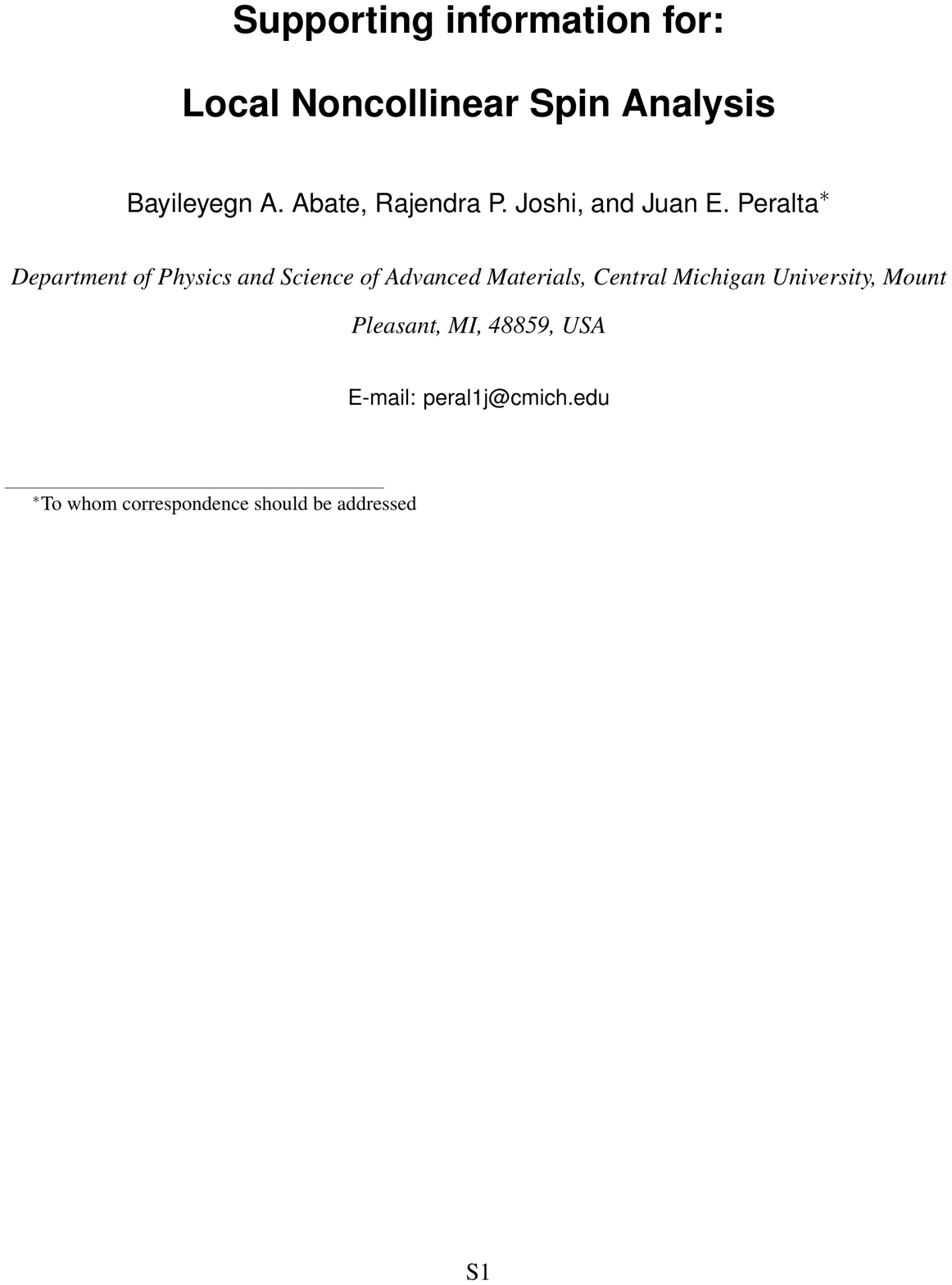}

\end{document}